# Exploring the potential of combining over- and under-stoichiometric MIEC materials for oxygen-ion batteries

S. Panisset[1,2], A. Schmid[3,*], A. Stangl[1,*], J. Fleig[3], D. Jauffres[2], M. Burriel[1,*]

1. Univ. Grenoble Alpes, CNRS, Grenoble INP, LMGP, 38000 Grenoble, France

2. Univ. Grenoble Alpes, CNRS, Grenoble INP, SIMaP, 38000 Grenoble, France

3. Institute of Chemical Technologies and Analytics, TU Wien, Vienna 1060, Austria

* alexander.e164.schmid@tuwien.ac.at, alexander.stangl@grenoble-inp.fr, monica.burriel@grenoble-inp.fr

## Abstract

The increasing demand for energy storage solutions has spurred intensive research into next-generation battery technologies. Oxygen-ion batteries (OIBs), which leverage mixed ionic-electronic conducting (MIEC) oxides, have emerged as promising candidates due to their solid, non-flammable nature and potential for high power densities. This study investigates the use of over-stoichiometric $La_2NiO_{4+\delta}$ (L2NO4) as a cathode material for OIBs, exploring its capacity for electrochemical energy storage. Half-cell measurements reveal that L2NO4 with a closed-pore microstructure can store oxygen, achieving a volumetric charge of 63 mA.h.cm$^{-3}$ at 400 °C with a current density of 3.6 µA.cm$^{-2}$ and potentials up to 0.75 V vs. 1 bar $O_2$. Additionally, a functional full cell combining over-stoichiometric L2NO4 and under-stoichiometric $La_{0.5}Sr_{0.5}Cr_{0.2}Mn_{0.8}O_{3-\delta}$ (LSCrMn) has been successfully developed, demonstrating excellent cyclability and coulomb efficiency. The full cell reaches a maximum volumetric charge of 90 mA.h.cm$^{-3}$ at 400 °C, 17.8 µA.cm$^{-2}$, and a cut-off voltage of 1.8 V. This proof of concept underscores the viability of combining over- and under-stoichiometric MIEC materials in OIBs and provides critical insights into optimizing electrode materials and tuning oxygen content for improved performance. This research lays the groundwork for future advancements in OIB technology, aiming to develop materials with lower resistance and higher efficiency.

## 1. Introduction

The increasing demand for energy storage solutions to power the growing network of portable electronics or electric vehicles through renewable energy systems has stimulated an intensive exploration of next-generation battery technologies. At the heart of research into efficient energy storage devices, a novel class of batteries has recently emerged [1]. Known as oxygen-ion batteries (OIBs), these devices are based on a solid oxide electrolyte sandwiched between two mixed ionic-electronic conducting (MIEC) oxides, originally derived from the fields of solid oxide fuel cells (SOFCs) [2], electrolysis cells (SOECs) [3], and oxygen separation membranes [4]. They rely on the MIEC materials' ability to conduct both electrons and oxygen ions, allowing for a variable oxygen stoichiometry. Well-known under-stoichiometric perovskites such as lanthanum strontium ferrite (LSF), lanthanum strontium cobalt (LSC) or lanthanum strontium cobalt ferrite (LSCF), can change their oxygen non-stoichiometry $\delta$ depending on the oxygen partial pressure or an applied voltage [5], [6], [7], [8]. Very similar to lithium-ion batteries (LIBs) or other post-lithium batteries (*e.g.* Na, Mg...) [9], in OIBs, the energy is stored chemically by changing the oxygen content in the electrodes by reduction or oxidation of the electrode material. Notably, these electrochemical redox reactions involve oxide ions from the electrolyte and electrons from the current collector, but do not include oxygen exchange with the atmosphere. Therefore, oxygen exchange with the atmosphere must be eliminated to prevent self-discharge and thus facilitate battery like behavior of the MIEC electrodes. The charging step consists of a transfer of oxygen (in the form of oxygen ions) through the electrolyte from the negative electrode to the positive one and of electron holes through an external circuit. During the discharge, the oxygen ions flow back to the positive electrode, i.e. the one with the lower reducibility, until the oxygen chemical potentials of both electrodes equilibrate.

OIBs are subject to several advantages when compared to other battery technologies such as lithium-ion or sodium-sulfur batteries. Comprised solely of solid, non-flammable oxides, they pose significantly lower security risks in the event of device failure. Moreover, there is the potential to replace critical elements (e.g. Li, Co or Ni) with less critical ones and abundant elements (such as Fe, Ca, Ti, Cr, and Mn) thanks to the large selection of MIEC oxides, and the possible use of promising high entropy oxides. Furthermore, unlike SOCs, which typically operate at high temperatures (600-1000 °C), OIBs can function efficiently in the moderate temperature range (250 to 400 °C), making them more suitable for a wider range of applications in harsh conditions, without the need for complex thermal management systems. Additionally, OIBs offer the potential for higher power densities and faster charge/discharge rates compared to LIBs, depending on operation temperature. Overall, the flexibility and sustainability of oxygen ion batteries make them an attractive alternative to LIBs and SOCs for large scale energy storage applications such as supply/demand balancing systems for renewable energy sources and other stationary applications, e.g. for grid stabilization.

Recently, first proof-of-concept devices were developed and tested using $La_{0.6}Sr_{0.4}FeO_{3-\delta}$ (LSF) and $La_{0.5}Sr_{0.5}Cr_{0.2}Mn_{0.8}O_{3-\delta}$ (LSCrMn) thin film electrodes, deposited by pulsed laser deposition (PLD) on yttria-stabilized zirconia (YSZ) single crystal substrates [1]. Electrode capacities up to 350 mAh cm$^{-3}$ were found in half cell measurements at 350 to 500 °C. In this study, we extend the range of possible OIB electrodes to a novel family of over-stoichiometric oxides. MIEC materials with layered Ruddlesden-Popper type structure typically exhibit an oxygen hyper-stoichiometry ($\delta$) and can accommodate a wide range of interstitial oxygen concentration between $\delta$ = 0 and up to $\delta_{max}$ = 0.25 [10]. Similar to the variable oxygen sub-stoichiometry of acceptor doped perovskites, this variable hyper-stoichiometry may be exploited for electrochemical energy storage. This effect is demonstrated here for the Ruddlesden-Popper oxide $La_2NiO_{4+\delta}$ (L2NO4). Half-cell measurements are performed to assess the specific capacity and half-cell potential of L2NO4. In addition to the fully dense films, nano-columnar films with closed porosity are also deposited, characterized and evaluated. In these, a secondary charge storage mechanism, namely the storage of high-pressure molecular oxygen inside pores, may also play a relevant role [11]. Lastly, the possibility of combining an over-stoichiometric (L2NO4) and an under-stochiometric (LSCrMn) MIEC material to form a functional OIB full cell with extended power storage is explored for the first time.

## 2. Principle of Operation

The concept of OIB is based on the (electro)chemical storage of electrical energy by charging the chemical capacitance $C_{chem}$ of a MIEC material. In MIEC oxides, two types of charge carriers are involved: the ionic carriers (oxygen vacancies or ions located in lattice or interstitial sites) and the electronic carriers (electrons or electron holes). The chemical capacitance is a capacitive property MIEC materials that reflects the ability of an oxide to change its oxygen stoichiometry in response to a change in the oxygen chemical potential or an applied voltage [12]. It is different from a dielectric capacitance in that here no net charges are stored. Rather, any electronic charges added to or removed from the MIEC electrode are counterbalanced by a corresponding change in ionic charges, resulting in zero net charge. This leads to the storage of a formally neutral species, which in this case is an oxygen atom. The area specific chemical capacitance can be expressed as [13]:

$$C_{chem} = \pm \frac{4F^2 L}{V_m} \frac{\partial \delta}{\partial \mu_O} \qquad \text{Eq. 1}$$

where $\delta$ is the oxygen nonstoichiometry, $\mu_O$ is the oxygen chemical potential of a single oxygen atom, $V_m$ is the molar volume of the material, $L$ is the electrode thickness, and $F$ is the Faraday constant. Eq. 1 shows that the chemical capacitance provides information on the defect chemistry of the material as $C_{chem}$ is directly linked to the variations in $\delta$, which represents the concentration of oxygen vacancies or interstitials in the MIEC material. Indeed, several studies measured $C_{chem}$ by electrochemical impedance spectroscopy (EIS) to obtain the oxygen vacancy concentration in perovskite materials [13], [14], [15].

The capability of MIEC materials to change their oxygen stoichiometry based on the oxygen chemical potential allows them not only to conduct both electrons and oxygen ions, but also to store oxygen within their structure. This occurs through the filling of either oxygen vacancy sites or interstitials sites, accompanied by the formation of electron holes or the annihilation of electrons. Thus, the oxidation states of the transition metal ions in the electrode materials change, undergoing oxidation or reduction. This process is similar to what occurs in a lithium-ion battery (LIB), but in an OIB, oxygen ions are inserted and removed instead of lithium ions.

The half-cell potential of an OIB cathode or anode can be determined from the material's reducibility. The reducibility of an oxide is defined as the ability to undergo a chemical reduction reaction, or more precisely, it relates to how easily the material can change its oxidation state by losing oxygen. Finally, to form an OIB, two materials with different half-cell potentials, i.e., with different reducibility, are combined with an electronically insulating, but oxygen ion conducting electrolyte between them.

The theoretical maximal electrode capacity can be estimated via the difference in oxygen non-stoichiometry ($\delta$) between the most oxidized and most reduced states of the electrode material, either as oxygen vacancies $[V_O^{\bullet\bullet}]$ or oxygen interstitials concentration, $[O_i'']$:

$$\Delta Q = 2e\Delta(\delta) \qquad \text{Eq. 2}$$

where $Q$ is the charge stored in the electrode and $e$ is the elementary charge. Then, by knowing the maximal oxygen non-stoichiometry range of a material, a theoretical half-cell capacity can be calculated. For MIEC materials that can accommodate both oxygen interstitials and vacancies, such as $La_{2-x}Sr_xNiO_{4\pm\delta}$ [16], the range is calculated using $\Delta\delta = \delta_{max} - \delta_{min}$ where $\delta_{max} > 0$ and $\delta_{min} < 0$. The delta range ($\Delta\delta$) found in the literature for typical MIEC materials include: $La_{0.6}Sr_{0.4}FeO_{3-\delta}$ (LSF) is around 0.2 [17], $La_{0.4}Sr_{0.6}CoO_{3-\delta}$ $\Delta\delta$ ~ 0.3 [18], $La_{0.5}Sr_{0.5}Cr_{0.2}Mn_{0.8}O_{3-\delta}$ $\Delta\delta$ ~ 0.25 [1], and L2NO4 $\Delta\delta$ ~ 0.25 [10]. The most common methods to measure $\delta$ in bulk samples include iodometric titration [19], [20], thermogravimetric analysis (TGA) [20], [21], and coulometric titration [5], [8]. Coulometric titration can also be used to measure $\delta$ in thin films [22].

During the charging process, an overpotential is applied to the electrode, which modifies the electrode effective oxygen partial pressure $p_{O_2,eff}$. The resulting oxygen partial pressure gradient drives the movement of oxygen ions through the electrolyte. Since oxygen ions cannot diffuse back in the absence of an external electronic circuit, this difference in oxygen chemical potential is maintained, ensuring long-term storage. The effective oxygen partial pressure of the WE, $p^{WE}_{O_2,eff}$, is given by the Nernst equation [23]:

$$p^{WE}_{O_2,eff} = p^{at}_{O_2} \times exp\left(\frac{4e\eta}{kT}\right) \qquad \text{Eq. 3}$$

where $p^{at}_{O_2}$ is the oxygen partial pressure in the surrounding atmosphere, $e$ is the elementary charge, $k$ is Boltzmann's constant, and $T$ is the temperature (in K). The cathode overpotential $\eta$, can be calculated from the voltage between the working (WE) and reference electrode (RE) $U_{DC}$ by correcting for ohmic losses in the electrolyte. In particular, when using an over-stoichiometric cathode, the cell voltage is increased during the charge, leading to incorporation of oxygen into the over-stoichiometric crystal structure via the interstitialcy mechanism [24], [25]. Eq. 3 shows that a positive overpotential leads to enormous effective oxygen partial pressures $p^{WE}_{O_2,eff}$ inside the film. For example, at $p^{at}_{O_2} = 10\ mbar$, a potential of 250 mV would result in an effective oxygen partial pressure inside the film $p^{WE}_{O_2,eff}$ of $3.10^5$ bar, which may cause serious damage to the capping layer.

While the half-cell capacity is limited by the variation of $\delta$, i.e. $\Delta\delta$, the full cell capacity depends additionally on the overlap of the $\Delta\delta$ range of the two electrodes with respect to the $p_{O_2,eff}$. This is illustrated in Fig. 1 a) and c), for two under-stoichiometric MIEC materials with different reducibility (but same $\Delta\delta$ range). As indicated in Fig. 1 a), the cell capacity, $C_{cell}$, is smaller than the half-cell capacities of the anode and the cathode ($C^{an}_{half}$, $C^{cath}_{half}$). To maximize the cell capacity, one needs to combine two electrode materials, where the changes in oxygen stoichiometry of each electrode take place in different pressure regions, with the ideal case - exhibiting no overlap - shown in Fig. 1 b) and d). This allows to exploit the whole $\Delta\delta$ range of both MIEC electrodes, and the full cell capacity approximates the half-cell capacity: $C_{cell} \rightarrow C_{half}$. So far, we have assumed that both electrodes show the same $\delta$ range. For materials with different $\Delta\delta$, one must adapt the electrode thicknesses, to guarantee that the total oxygen storage capacity is the same in both electrodes. Moreover, the difference of these pressure ranges, i.e. the difference in half cell potentials is what defines the cell voltage of the OIB. It is thus desirable to combine a material with a very low transition pressure and one with a very high one. The latter is difficult to achieve with under-stoichiometric materials but could be achieved with over-stoichiometric MIEC materials.

A second remark concerns the equilibrium pressure inside the two electrodes, which corresponds to the fully discharged state of the OIB and depends on the initial oxygen concentration of the two electrodes. Excess or depletion in oxygen will cause a deviation from the ideal equilibrium pressure and will result in a reduced cell capacity. This can be avoided by controlling the initial oxygen concentration via a third reference electrode, as discussed in detail below.

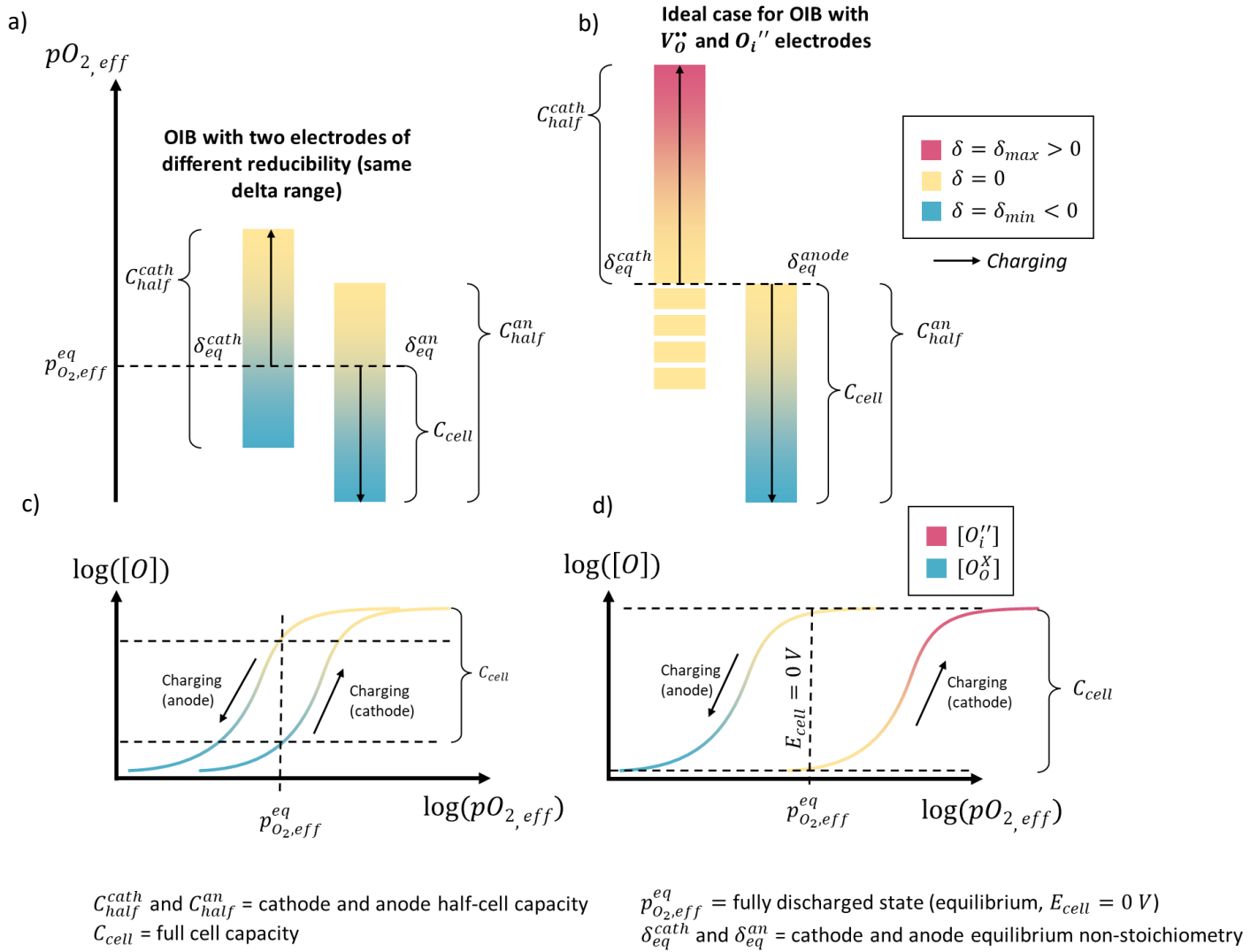

$C_{half}^{cath}$ and $C_{half}^{an}$ = cathode and anode half-cell capacity
$C_{cell}$ = full cell capacity

$p_{O_2,eff}^{eq}$ = fully discharged state (equilibrium, $E_{cell} = 0\ V$)
$\delta_{eq}^{cath}$ and $\delta_{eq}^{an}$ = cathode and anode equilibrium non-stoichiometry

Fig. 1: Oxygen non-stoichiometry range and half-cell capacity of a) two under-stoichiometric MIEC with different reducibility and same $\delta$ range, and b) the combination of one over-stoichiometric MIEC ($O_i''$) for the cathode and an under-stoichiometric MIEC ($V_O^{\bullet\bullet}$) for the anode. c) and d) represent schematically the Brouwer diagram for the case a) and b), respectively. The black dashed lines correspond to the equilibrium of the two half-cell electrodes (fully discharged state). The black arrows show the charging direction, meaning the cathode oxygen filling and the anode oxygen removal. [O] in the Brouwer diagram (c and d) refers to the normalized oxygen concentration.

# 3. Material and Methods

## 3.1 Sample preparation

Lanthanum nickel oxide La$_2$NiO$_{4+\delta}$ (L2NO4) working electrode (WE) thin films were grown using Pulsed Injection Metal Organic Chemical Vapor Deposition (PI-MOCVD) on 1 x 1 cm² polished Yttria-Stabilized Zirconia (YSZ) (100 oriented) single crystals (Crystec Gmbh). An alumina mask placed on top of the substrate is used to deposit a 0.75 x 0.75 cm² electrode. The precursor solution was prepared by dissolving La(TMHD)$_3$ and Ni(TMHD)$_2$ (TMHD = 2,2,6,6-tetramethylheptane-3,5-dionate), both from Strem Chemicals, in m-xylene (1,3-dimethylbenzene) from Alfa Aesar at a concentration of 0.02 mol/L. The injection frequency and opening times were set at 4 Hz and 2 ms, respectively. The carrier gas concentration was 34% Ar and 66% O$_2$ and the total pressure inside the reactor was kept at 5 Torr. The ratio La/Ni in solution has been adapted to the deposition temperature to consider the difference of reaction kinetics of the film growth at different temperatures. The ratio is set to 4 and 3.15 at 600 °C and 750 °C, respectively. At lower temperatures (600 °C), the film exhibits a nanocolumnar morphology, while at 750 °C, it grows dense [33]. To obtain closed-pores as shown Fig. 3 a) and b), the process involves two consecutive deposition steps. Initially, a 200 nm layer of porous L2NO4 is grown at 600 °C, followed by L2NO4 capping layer at 750 °C.

Next, a 100 nm dense platinum (Pt) layer is deposited using the evaporator MEB550 from Plassys at a deposition rate of 0.5 nm.s$^{-1}$. The whole surface of the cell is covered by the Pt current collector (CC) as shown in Fig. 3. On top of the CC, a 100 nm dense silicon nitride (Si$_3$N$_4$) oxygen blocking layer is deposited by Plasma Enhanced-CVD as capping layer. Two stripes of 1 mm width at the edge of the cell are masked during the capping to ensure electrical contact to the CC below.

As illustrated in Fig. 2, a dense 100 nm films of LSF and LSCrMn were deposited by Pulsed Laser Deposition (PLD) to serve as the counter and reference electrodes (CE and RE) for the half-cell and the full cell, respectively. On top of the counter and reference electrodes, a 100 nm dense Pt layer is then sputtered with a rate of 0.5 nm.s$^{-1}$. For the full cell, a capping layer is added on the counter electrode, which consists of a 200 nm dense ZrO$_2$ layer deposited by PLD. PLD depositions were done using a KrF excimer laser (248 nm) with a pulse energy of 100 mJ (fluence ca. 1 J.cm$^{-2}$) and a pulse rate of 10 Hz, and the target to substrate distance was 6 cm for all depositions. Additional deposition parameters are listed in Table 1.

*Table 1: PLD deposition parameters*

| Material | Substrate temperature | Oxygen pressure | Growth Rate |
|---|---|---|---|
| LSF | 600 °C | 4 Pa | 14 nm/pulse |
| *LSCrMn* | 700 °C | 1.5 Pa | 18 nm/pulse |
| *ZrO₂*  | 650 °C | 1.5 Pa | 16 nm/pulse |

### 3.2 Electrochemical characterization

Galvanostatic cycling measurements have been performed on both half cells and full cell samples using a Keithley 2600 source meter in galvanostatic mode in a 4-wire configuration, with currents from 1 to 20 µA and in a temperature range of 350 to 400 °C. Both types of cells were cycled several times between a lower and upper cutoff voltage depending on the temperature, current and cell configuration. The voltage at the working electrode (WE) was recorded against the reference electrode open to (and thus in equilibrium with) the atmosphere an oxygen partial pressure of 10 mbar. Subsequently, the potential against 1 bar of oxygen was computed using Nernst's equation. Furthermore, the cell voltage between the working and counter electrodes was measured with a separate voltmeter (Keithley 2000).

## 4. Results

### 4.1 $La_2NiO_{4+\delta}$ working electrode microstructure

A schematic representation of the half-cell and full cell architectures is depicted in Fig. 2. $La_2NiO_{4+\delta}$ (L2NO4) thin films were deposited by Pulsed Injection Metal Organic Chemical Vapor Deposition (PI-MOCVD) on yttria-stabilized zirconia (YSZ) substrates. Dense 100 nm films of LSF and LSCrMn were utilized as the counter (CE) and reference electrodes (RE) and covered by sputtered platinum for the current collection. $Si_3N_4$ oxygen blocking layers were used to prevent the oxygen exchange of the WE with the atmosphere, and, for the full cell only, the CE is capped with a $ZrO_2$ blocking layer. In this study, three distinct $La_2NiO_{4+\delta}$ (L2NO4) microstructures were fabricated and evaluated in half-cell configuration: a dense microstructure, a nano-columnar structure, and a closed-pore microstructure. To obtain these distinct morphologies the films were deposited at different temperatures: 750 °C for the dense structure, 600 °C for the nano-columnar structure, and a combination of 600 °C followed by 750°C for the closed-pore microstructure. Fig. 3 a) and b) show the cross-section of a L2NO4 dense and nano-columnar film obtained at 750 °C and 600 °C, respectively, while Fig. 3 c) and d) corresponds to the film with closed-pore. The SEM and TEM cross-section images shown in Fig. 3 c) and d) reveal the presence of closed-pores in the bottom 200 nm of the L2NO4 film while the top 400 nm of the film are totally dense. Even at low deposition temperatures, the beneficial formation of a thin, dense bottom interface L2NO4 layer is observed, ensuring good adhesion and ionic contact of the film with the YSZ electrolyte. Above the porous part of the film, 3 times larger grains (130 $\pm$ 40 nm width) constitute the dense part obtained at the high deposition temperature (750 °C). A SEM top view of the dense surface is depicted in Fig. S2 .

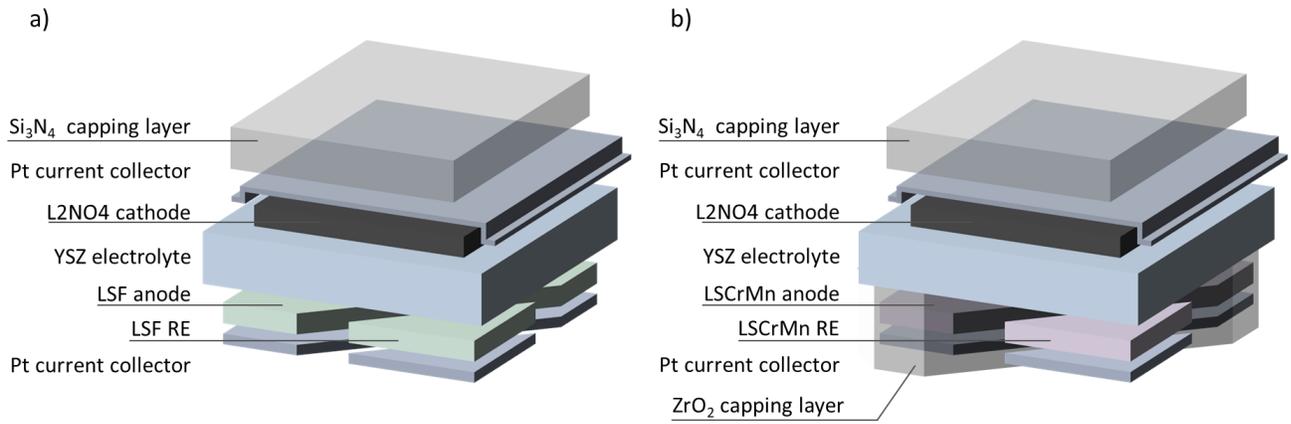

Fig. 2: Schematic representation of a a) OIB half-cell and b) full cell. The main difference between the two devices is the capping layer on the anode side. RE is the reference electrode.

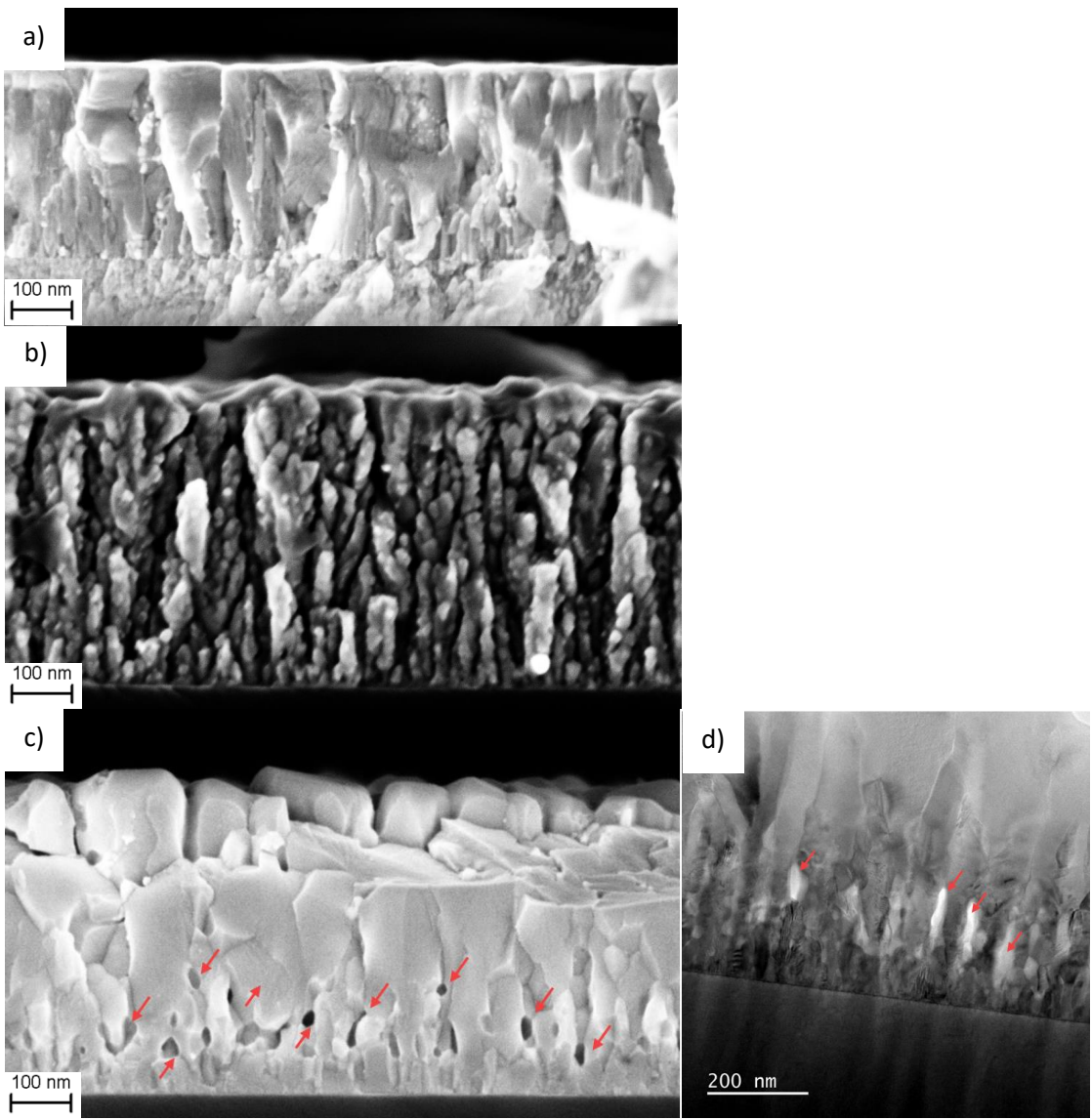

*Fig. 3: Cross-sectional SEM view of L2NO4 electrode with a) dense, b) nano-columnar, and c) closed-pore microstructure. TEM view of the closed-pore microstructure (d) at the L2NO4/YSZ interface. The red arrows indicate closed-pore.*

### 4.2  L2NO4/YSZ/LSF half-cells

L2NO4 belongs to the first order Ruddlesden-Popper family of materials with the $A_2BO_4$ structure. The structure of L2NO4 can be described as alternating layers of the perovskite ($LaNiO_3$) and rock salt (LaO) structures along the out-of-plane crystallographic direction. L2NO4 can accommodate an excess of oxygen $\delta$ on the interstitial sites in the rock salt layer LaO, typically around $\delta = 0.16$ at room temperature and atmospheric pressure [26], [27]. This leads to the partial oxidation of $Ni^{2+}$ to $Ni^{3+}$. When the exchange with atmosphere is prevented, the overall reaction in the L2NO4 half-cell is defined as:

$$La_2NiO_{4+\delta} + x\, O_{YSZ}^{2-} \leftrightarrows La_2NiO_{4+\delta+x} + 2x\, e_{cc}^{-} \qquad \text{Eq. 4}$$

As a first step, the L2NO4 electrode is brought into equilibrium with the atmosphere by short-circuiting the WE with the CE and the reference electrode, the latter being open to atmosphere ($pO_2$ = 10 mbar). This enables the oxygen exchange between the WE, CE and RE and thus also with the controlled atmosphere through the electrolyte. The initially oxygen over-stoichiometric electrode material equilibrates with the oxygen partial pressure. The galvanostatic cycling measurements can start when the OCV is close to 0 V vs RE.

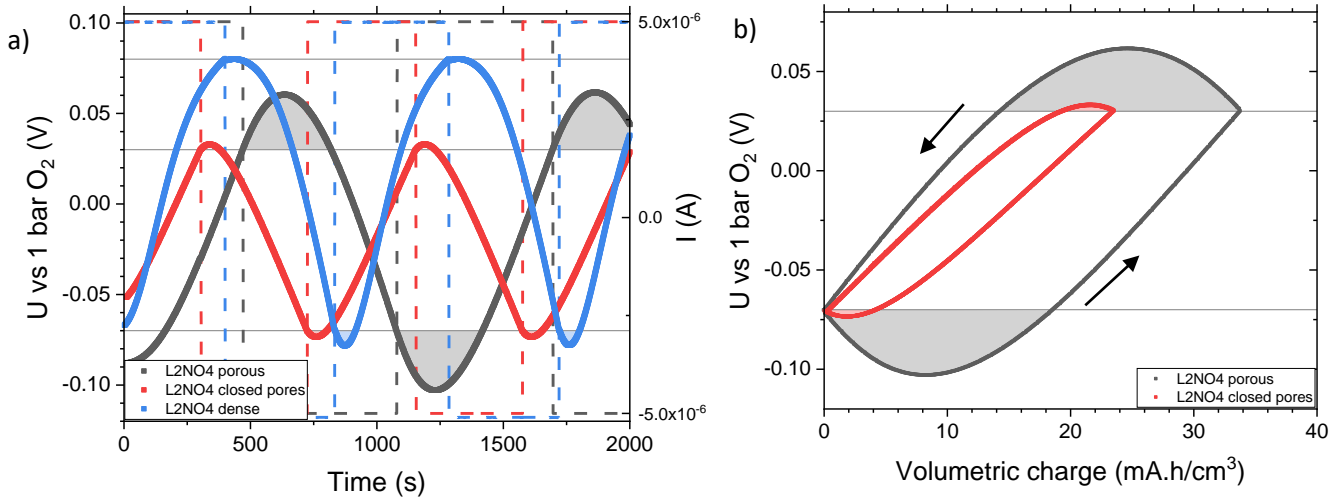

*Fig. 4: Galvanostatic cycling measurement of a half-cell at 400 °C and at 1.8 µA.cm$^{-2}$ for 3 different electrode microstructures (dense, nano-columnar, closed-pore). a) Temporal evolution of the WE potential vs RE and the applied current (dashed lines), b) "reversed" charge/discharge cycles for porous (black) and closed-pore (red) L2NO4 WE. The dense electrode has been cycled up to 0.15 V vs RE, whereas the porous and closed-pore electrodes were cycled up to 0.1 V vs RE. The colored zones are visual aids to see where the cutoff voltages have been exceeded. The volumetric charge is normalized by the WE volume. The first cycle is not shown.*

Fig. 4 a) shows the temporal variation of the WE potential against the reference electrode for the three L2NO4 film microstructures studied. The upper cutoff voltage was set to 100 mV, for the porous and the closed-pore electrodes and 150 mV for the dense electrode. The increase (decrease) of the potential difference is related to the incorporation (excorporation) of oxygen into the L2NO4 electrode through the CE, and thus charging and discharging, respectively. An unexpected voltage overshoot above the 100 mV limit was measured at the beginning of the discharge step for the porous sample, while it was less pronounced in the closed-pore sample. The same phenomenon is observed below the lower limit at the beginning of the charging step as highlighted by the colored regions in Fig. 4. Even after switching the current (discharge → charge) as indicated by the vertical dashed lines, the potential does not increase immediately leading to the unexpected behavior where the measured electrode potential continues to decrease even during the beginning of the charging step.

As can be observed in Fig. 4, the overshoot depends on the microstructure of the L2NO4 and is accentuated by the presence of pores, the extreme case being the nanocolumnar L2NO4 electrode. This behavior correlates with the effective in-plane conductivity of the different film morphologies. For nano-columnar films, the effective in-plane ionic conductivity is much lower compared to the two other microstructures as the nano-columns are only connected at the bottom of the electrode. Correspondingly, we observe a strong overshoot during the charging and discharging steps. The voltage overshoot is significantly reduced in the case of the closed-pore microstructure, probably due to the presence of the upper dense layer improving the in-plane conductivity and is nearly absent for the dense electrode (cutoff voltage was set to 0.15 V for this measurement).

We believe that this overshoot is most likely a measurement artefact, caused by the formation of a lateral oxygen concentration gradient inside the WE favored by the asymmetrical placement of the counter and reference electrodes as illustrated in Fig. S3. A more detailed explanation is provided in the Supplementary Information. Indeed, the counter electrode is not perfectly aligned with the whole WE as the reference has been placed on one corner of the substrate (see Fig. 2). Asymmetrical electrode placement has been reported to also cause artifacts in impedance spectroscopy [28].

Fig. 5 shows the WE potential versus the CE. Notably, the overshoot artifact due to the film inhomogeneity does not appear. In counterpart, we cannot distinguish between WE and CE potentials, i.e., the measured potential difference includes the overpotentials due to transport resistances in both electrodes and in the electrolyte and the difference in OCV of both electrodes, i.e.:

$$U = OCV_{WE} - OCV_{CE} + I\ (R_{YSZ} + R_{WE} + R_{CE}) \qquad \text{Eq. 5}$$

where *I* is the current, defined as positive in charging direction, and the R terms denote the transport resistances in the respective components, including possible charge transfer resistances. We may now assume that the voltage drops due to transport are equal in both current directions, and further, that the counter electrode equilibrates sufficiently fast with the atmosphere, such that its OCV with respect to that atmosphere is always zero. Then, by averaging the voltages measured in both charging and discharging directions at equal state of charge, because the charge and discharge current are equal but opposite in sign, we obtain:

$$0.5\ (\vec{U} + \overleftarrow{U}) = 0.5\ (OCV_{WE} - OCV_{CE} + \vec{I}\,R + OCV_{WE} - OCV_{CE} + \overleftarrow{I}R) = OCV_{WE} \qquad \text{Eq. 6}$$

where $R$ is the total cell internal resistance ($R = R_{YSZ} + R_{WE} + R_{CE}$) and the arrows indicate charging or discharging.

The resulting OCV curves are shown in Fig. 5 as dashed lines. The fact that OCV starts very close to 0 mV supports this estimation, confirming that the counter electrode equilibrates quickly enough with the atmosphere to maintain a near-zero OCV. Very similar OCV curves have been obtained for increasing upper cutoff voltages (green, blue and red curves in Fig. 5) until 250 mV, where a strong deviation occurs above 0.8 V (grey curve). This may be a sign of degradation of the CE (LSF decomposition), or a break in the sealing layer occurring at high voltages. Correspondingly, at the end of the discharge, not all the charges have been restored, i.e., the Coulomb efficiency is distinctly below unity. Considering that such high potentials correspond to effective pressures (i.e. fugacities) orders of magnitude above atmospheric pressure (typically $10^4$ bar at 200 mV and $10^5$ bar at 250 mV), mechanical failure of the sealing layer is highly plausible. Fig. S5 illustrates serious damages of the sealing layer of the L2NO4 half-cell after galvanostatic cycling measurements at 400 °C, 250 mV and 8.9 µA.cm$^{-2}$. In the tested conditions, the mechanism of gaseous oxygen formation into the closed-pores has not been detected. Krammer *et al.* showed that above a critical potential, this additional storage mechanism becomes relevant and leads to a slope decrease in the charge curve [11]. Here, the observed slope discontinuity is attributed to the loss of charges as they are not restored during the discharge. However, we cannot exclude that this mechanism also contributes to the oxygen storage. We conclude that 0.2 V vs RE (0.13 V vs 1 bar O$_2$) is safe to cycle the cell in a reproducible way without damaging the WE.

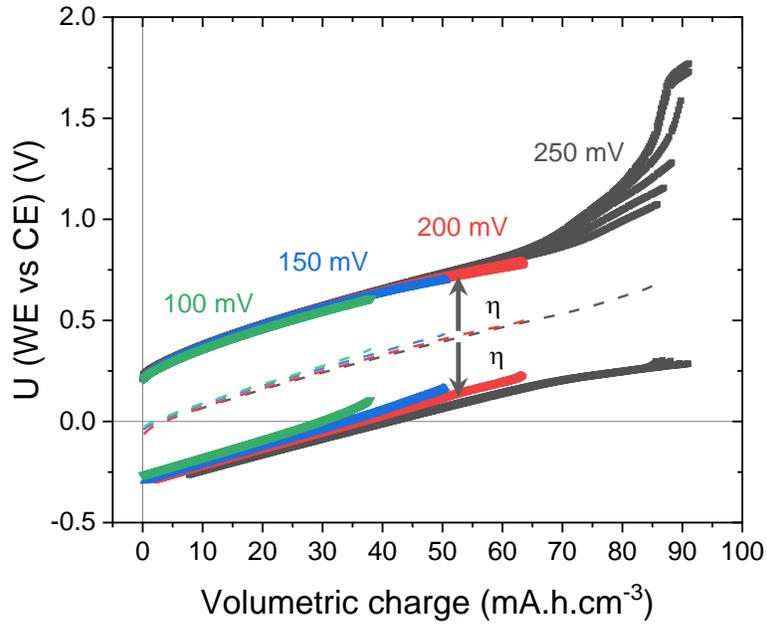

*Fig. 5: Half-cell potential of closed-pore WE versus CE during charge and discharge for various cutoff voltages from 100 to 250 mV (set between WE and RE). The dashed lines correspond to the OCV against the volumetric charge. The volumetric charge is calculated using the WE volume. Up to 200 mV vs RE, the half-cell can be cycled safely.*

In literature, the bulk phase diagram of L2NO4 indicates that in the orthorhombic phase (space group Fmmm) the oxygen over-stoichiometric $\delta$ can reach a value of up to 0.2 [29]. The maximal $\delta$ = 0.25 has been reported with the existence of a monoclinic superstructure (space group $C_2$) at room temperature [30]. By considering that the L2NO4 can be completely reduced ($\delta$ = 0), the L2NO4 maximum theoretical capacity is equal to 235 mA.h.cm$^{-3}$. For comparison the maximal theoretical capacity of LSCrMn, the highest among the materials tested so far, is equal to 385 mA.h.cm$^{-3}$ [1]. In this study, a much lower capacity has been obtained using closed-pore microstructure (63 mA.h.cm$^{-3}$ at 400 °C). With an oxygen partial pressure of 10 mbar, the oxygen non-stochiometric of L2NO4 is about $\delta \approx 0.09$ according to previous work [32], which means that the electrode is not fully empty at equilibrium (discharged state). Based on the measured volume specific capacity (63 mA.h.cm$^{-3}$), we can estimate a $\Delta\delta \approx 0.07$ leading to a $\delta_{charged} \approx 0.16$ for an effective partial pressure $p_{O_2,eff} = 1.10^4$ bar, which is still relatively far from the $\delta_{max} \sim 0.25$. Altogether, these half-cell measurements demonstrate that charge can be reversibly stored in L2NO4, with a volume specific capacity of 63 mA.h.cm$^{-3}$ at potentials up to 0.75 V vs. 1 bar $O_2$. We expect that these values can be increased by improving the capping layer.

### 4.3  L2NO4/YSZ/LSCrMn full cell

In the following, we demonstrate the successful operation of L2NO4 with closed-pore microstructure as cathode in OIB full cells. The previously studied LSCrMn electrode is used as the anode to benefit from the low reducibility and high capacity of this material [1]. The terms *cathode* and *anode* refer to the electrode reduction and oxidation, respectively, during discharging. In this battery, the oxygen ions are expected to fill empty interstitial sites of the L2NO4 cathode during charging, while oxygen vacancies are formed in the LSCrMn anode. During discharging, the oxygen vacancies of LSCrMn anode are expected to be filled with the interstitial oxygen ions coming from the L2NO4 cathode, which is reduced.

Before any measurements, the cell is short-circuited to equilibrate both electrodes with the atmosphere ($pO_2$ = 10 mbar). During this step, both electrode potentials are equilibrated by exchanging oxygen with the atmosphere through the reference electrode. The full cell has been tested over the temperature range of 350-450 °C at various cutoff voltages and applied current densities.

Fig. 6 shows that at 400 °C and for a cutoff voltage of 1.4 V, the cell reaches a maximum capacity of 103 mA.h.cm$^{-3}$ at 1.8 µA.cm$^{-2}$ but of only 23.4 mA.h.cm$^{-3}$ at 17.8 µA.cm$^{-2}$. The cell successfully charges and discharges, demonstrating the good functioning of the capping layers. During charging, the cell voltage under load is the sum of the OCV difference of both electrodes and the kinetic overpotentials in the electrodes and electrolyte. At higher current densities, kinetic limitations of the oxygen ion diffusion in the electrodes and the electrolyte increase their resistive contribution during the charging process. Thus, the cut-off voltage is already reached at a lower oxygen over-stoichiometry, limiting the capacity resulting in a lower OCV difference, i.e. in less utilization of the storage capacity. For the lowest current density (1.8 µA.cm$^{-2}$) we also observe a coulomb efficiency distinctly less than unity (0.82), meaning that a portion of the oxygen inserted into the electrode is not retrieved afterwards. Specifically, about 100 mA.h.cm$^{-3}$ are inserted into the electrode but only about 80 are retrieved. Interestingly, during charging, at the same point, at 80 mA.h.cm$^{-3}$, the curve gets markedly flatter. This may hint different mechanisms being responsible for the reversible storage and the irreversible excess during charging. One possible explanation is a leakage of oxygen into the atmosphere either through damages in the capping layer or the platinum current collector of the WE. At higher current density, this phenomenon is not visible, likely because the WE does not excess the critical potential to reach the upper cutoff voltage imposed during the cycling test. However, even for high current densities (17.8 µA.cm$^{-2}$) if the cutoff voltage is increased, for instance to 1.8 V (see Fig. 6 b), the change of slope and a distinct coulomb efficiency lower than unity are again observed. Also, the higher cut-off voltage allows for a higher utilization of the electrode, reaching a maximum capacity of 122.8 mA.h.cm$^{-3}$ and gradually stabilizes around 90 mA.h.cm$^{-3}$.

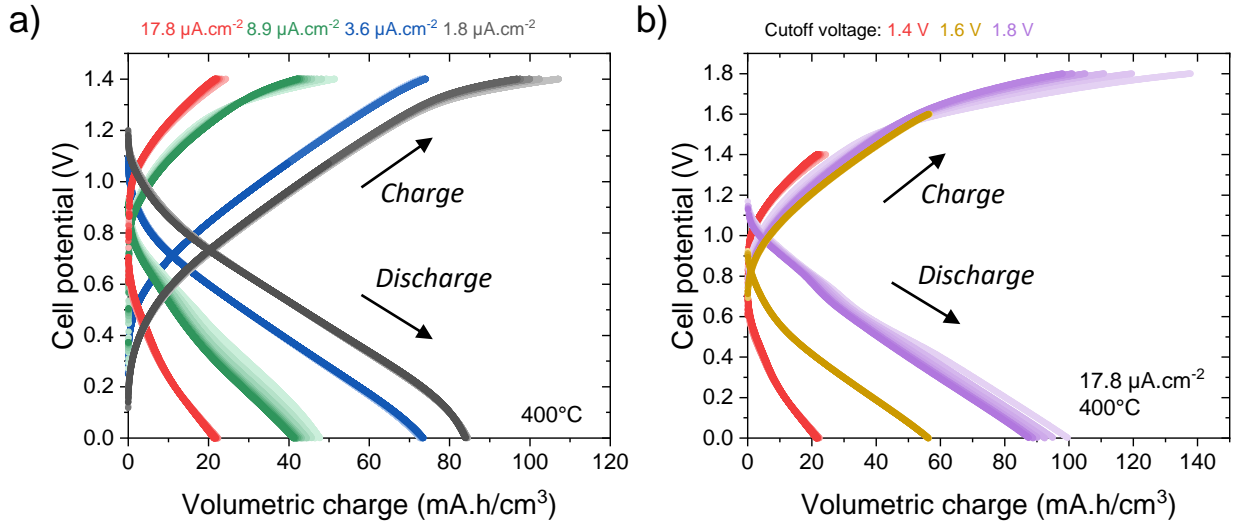

Fig. 6: Full cell charge-discharge curves with L2NO4 as positive electrode and LSCrMn as negative electrode. a) Cell potential at 400 °C for different current densities ranging from 1.8 to 17.8 µA.cm$^{-2}$ for a cutoff voltage of 1.4 V, and b) cell potential at 400 °C and 17.8 µA.cm$^{-2}$ for different cutoff voltages ranging from 1.4 to 1.8 V. The first cycle has been removed.

Table 2 summarizes the maximal capacity normalized by both electrode volume and the related coulomb efficiency over the 10 first cycles obtained for the L2NO4/YSZ/LSCrMn full cell at various temperature, current density and upper cutoff voltages. As for the half cell, the first charging step is removed. In most of the cases, the coulomb efficiency is close to unity. The full cell showed a stable capacity of about 90 mA.h.cm$^{-3}$ at 400°C, 17.8 µA.cm$^{-2}$ and a cutoff voltage of 1.8 V, a value slightly lower than what has been achieved before in similar conditions with LSF cathode [1]. Surprisingly, this capacity exceeds all the values measured in the half cell samples using L2NO4 as cathode. This may be a consequence of a better sealing layer, allowing the positive L2NO4 electrode to go to higher potentials, and thus enabling higher utilization of its theoretical capacity. Another possible effect is that, due to parasitic oxygen release into the atmosphere, oxygen may be removed from the full cell, which would lead to a situation where during the discharge step the L2NO4 electrode potential is decreased below its initial state (estimated to be about $\delta \approx 0.09$, equilibrated to 10 mbar) which in turn would make accessible a larger $\Delta\delta$ window. As prospects, the initialization process can be optimized to obtain a larger $\Delta\delta$ window as explained in detail in the Supplementary Information.

Table 2: Summary of maximum capacity and coulomb efficiency values of the L2NO4/YSZ/LSCrMn full cell obtained at various temperature, current density and upper cutoff voltages

| Temperature (°C) | Current density (µA.cm$^{-2}$) | Cutoff voltage (V vs CE) | | Max capacity (mA.h.cm$^{-3}$) | Coulomb efficiency (%) |
|---|---|---|---|---|---|
| | | Lower | Upper | | |
| 350 | 8.9 | 0 | 1.8 | 47.5 | 93.0 |
| 350 | 17.8 | 0 | 1.8 | 6.1 | 92.6 |
| 400 | 1.8 | 0 | 1.4 | 103.2 | 81.7 |
| 400 | 3.6 | 0 | 1.4 | 73.4 | 99.5 |
| 400 | 8.9 | 0 | 1.4 | 49.7 | 93.8 |
| 400 | 17.8 | 0 | 1.4 | 23.4 | 94.4 |
| 400 | 17.8 | 0 | 1.6 | 56.6 | 99.4 |
| 400 | 17.8 | 0 | 1.8 | 122.8 | 77.9 |
| 450 | 3.6 | 0 | 1.3 | 104.5 | 78.2 |
| 450 | 8.9 | 0 | 1.4 | 92.9 | 87.9 |
| 450 | 17.8 | 0 | 1.3 | 51.5 | 99.9 |
| 450 | 17.8 | 0 | 1.4 | 60.5 | 99.9 |

## 5. Conclusions

This study demonstrates the utilization of over-stoichiometric MIEC materials as OIB cathode. It showed that lanthanum nickel oxide ($La_2NiO_{4+\delta}$) half-cell with closed-pore microstructure can store over-stoichiometric oxygen reaching a volumetric charge of 63 mA.h.cm$^{-3}$ at 400°C, using a current density of 3.6 µA.cm$^{-2}$, and at potentials up to 0.75 V vs. 1 bar oxygen. A full cell has been successfully manufactured using $La_2NiO_{4+\delta}$ for the first time, demonstrating that over- and under-stochiometric MIEC materials can be combined as cathode and anode, respectively, to form a functional OIB. We proved that, after equilibration with a controlled atmosphere and no pre-conditioning step to change the oxygen content from the electrodes, the full cell can be cycled with a very good cyclability and coulomb efficiency. A maximum volumetric charge of 90 mA.h.cm$^{-3}$ has been reached at 400°C, 17.8 µA.cm$^{-2}$, and a cutoff voltage of 1.8V. This study has achieved the proof of concept of the combination of over- and under-stoichiometric electrodes and contributes significantly to a better understanding of the OIB concept. The research provides valuable insights into how to create reliable OIBs, identifying the ideal conditions for choosing electrode materials, and understanding how to tune the oxygen content to utilize the full range of oxygen non-stoichiometry. Finally, this work lays the foundations for future studies aimed at developing materials with lower resistance and higher performance, thereby advancing the practical implementation of OIB technology.

## Author Contributions

Conceptualization: ASc, ASt, MB, DJ, JF; Data curation: SP, ASc; Methodology: ASc, ASt; Formal analysis: SP; Investigation: SP, ASc; Validation: ASc; Visualization: ASt; Funding acquisition: DJ, MB; Supervision: DJ, MB; Writing – original draft: SP; Writing – review & editing: SP, ASc, ASt, MB, DJ, JF.

## Acknowledgements

This work was funded by the European Union's Horizon 2020 research and innovation program under grant agreements no. 824072 (Harvestore project) and by the Centre of Excellence of Multifunctional Architectured Materials "CEMAM" n°ANR-10-LABX-44-01 as part of the "Investments for the Future" Program. This research has benefited from characterization equipment of the Grenoble INP - CMTC platform.

## Data Availability Statement

The data that support the findings of this study are openly available in Zenodo at https://zenodo.org/12528114, reference number 10.5281/zenodo.12528114

## Conflicts of interest

There are no conflicts to declare.

# Supporting information

# Exploring the potential of combining over- and under-stoichiometric MIEC materials for oxygen-ion batteries

S. Panisset[1,2], A. Schmid[3,*], A. Stangl[1,*], J. Fleig[3], D. Jauffres[2], M. Burriel[1,*]

1. Univ. Grenoble Alpes, CNRS, Grenoble INP, LMGP, 38000 Grenoble, France

2. Univ. Grenoble Alpes, CNRS, Grenoble INP, SIMaP, 38000 Grenoble, France

3. Institute of Chemical Technologies and Analytics, TU Wien, Vienna 1060, Austria

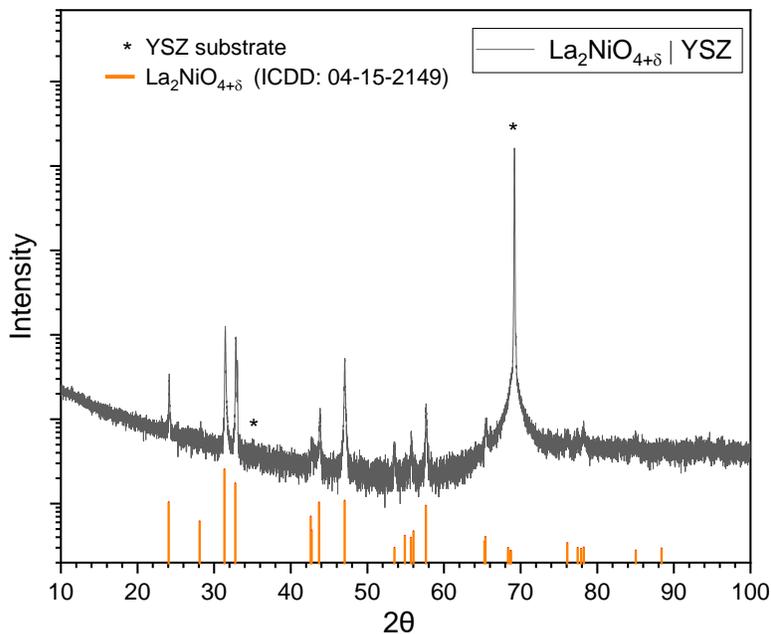

*Fig. S1: XRD of closed-pores L2NO4 electrode used for half-cell and full-cell samples*

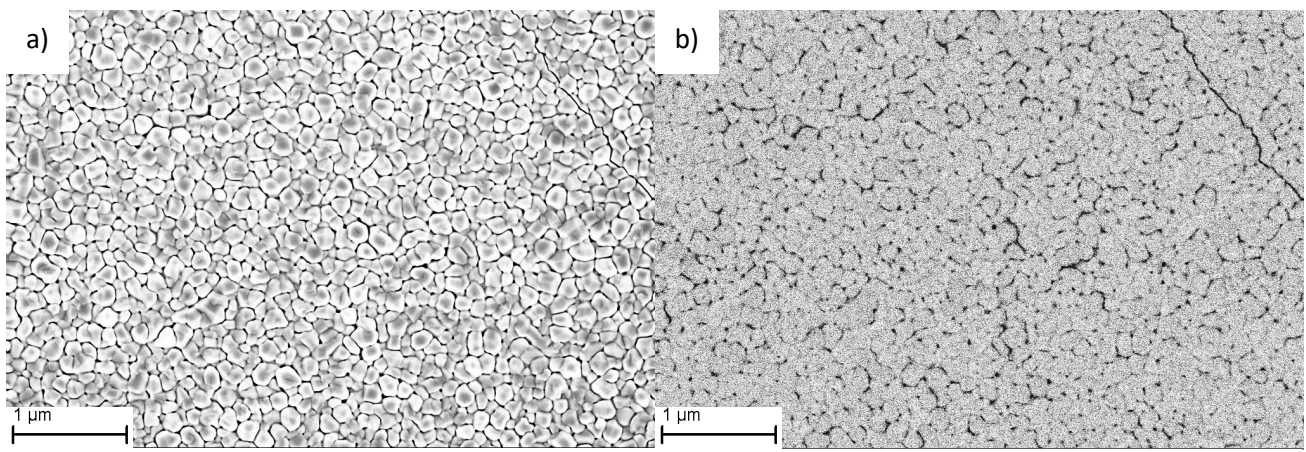

*Fig. S2: SEM top view of the L2NO4 electrode with the closed pores microstructure in secondary electron (SE) (a) and in Electron backscattered (EBS) (b)*

**Half-cell measurement artefacts**

We believe that this overshoot is most likely a measurement artefact, caused by the formation of a lateral oxygen concentration gradient inside the WE favored by the asymmetrical placement of the counter and reference electrodes as illustrated in Fig. S3. Indeed, the counter electrode is not perfectly aligned with the whole WE as the reference has been placed on one corner of the substrate (see Fig. 2). Asymmetrical electrode placement has been reported to also cause artifacts in impedance spectroscopy [1]. Furthermore, this geometry may also cause a non-negligible oxygen concentration gradient in the WE, due to the limited ionic in-plane conductivity of the L2NO4 film. This results in an inhomogeneity of the oxygen concentration in the L2NO4 material that causes a higher oxygen concentration just above the CE, and thus, an in-plane ionic diffusion (produced by the oxygen concentration gradient in the film).

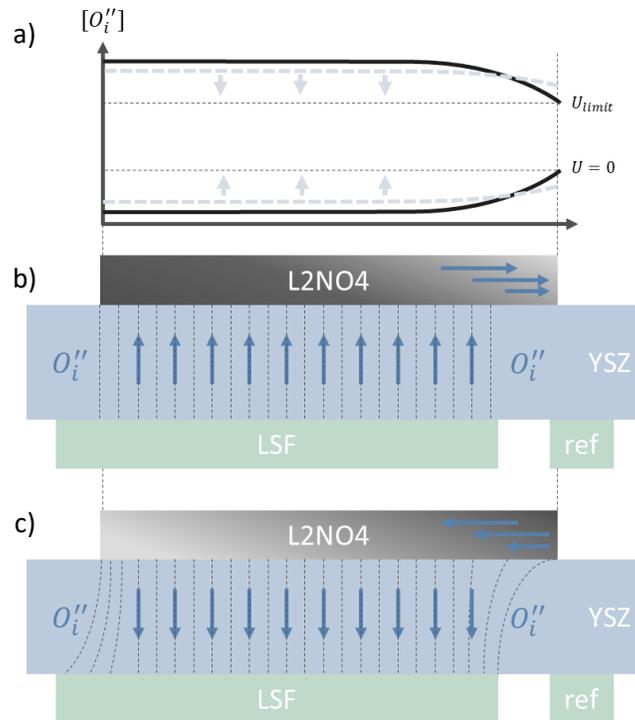

*Fig. S3: Schematic representation of the oxygen concentration gradient in the L2NO4 WE: a) oxygen concentration level evolution across the electrode length just after switching the current (black lines for charge → discharge and discharge → charge). In grey, the oxygen concentration level is uniformized by oxygen in-plane conduction. b) and c) shows the non-uniform oxygen incorporation and excorporation inside the WE during charging and discharging, respectively.*

This interpretation is consistent with the behavior observed when the temperature and the current density are changed. As exemplified in Fig. S4 a), the potential overshoot increases with the applied current and with the decrease in temperature. On the one hand, a higher current density implies a faster incorporation of oxygen in the L2NO4 film that may cause a higher oxygen concentration gradient. On the other hand, a lower temperature (Fig. S4 b) results in a lower ionic conductivity for L2NO4, which also contributes to the increase of the oxygen concentration gradient.

Altogether, it is thus highly plausible, that the limited ionic conductivity and the asymmetrical electrode configuration led to a distortion of the current density distribution in the electrolyte depending on the electrode state of charge and current direction, and thus ultimately changes in the equipotential line that is picked up by the reference electrode. However, a complete, mechanistic resolution and description of this effect is beyond the scope of this work. Instead, we may instead use a different approach described in the following.

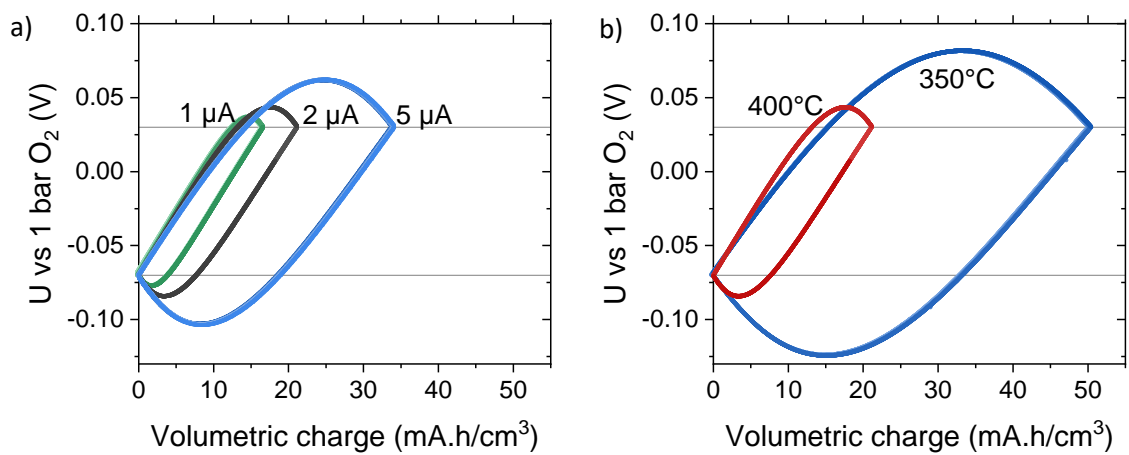

*Fig. S4: Galvanostatic cycling measurements of L2NO4 half-cell with porous WE at 10 mbar of $O_2$ for various conditions: a) applied current density at 400°C for a porous electrode and b) temperature at 8.9 µA.cm$^{-2}$.*

**Post-mortem sealing layer analysis**

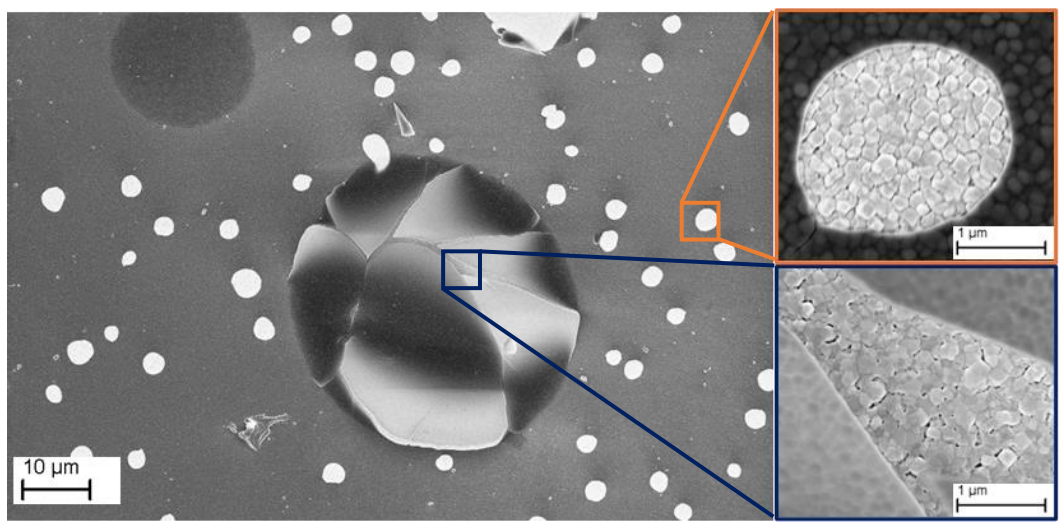

*Fig. S5: SEM pictures and magnified regions where the damaged $Si_3N_4$ capping layer capping layer can be observed, revealing the L2NO4 electrode film underneath.*

Two different types of damage can be observed. First, at the center, the sealing layer would have exploded revealing the L2NO4 electrode through the cracks (see the bottom right magnification SEM image). It seems that a local increase of the pressure could cause the mechanical failure of the sealing layer. Secondly, small holes (1 to 2 µm of diameter) were formed over the entire surface of the cell, again leaving the L2NO4 electrode in contact with the atmosphere.

## Initialization process and optimization of the Δδ window

In order to make accessible a larger Δδ window, we propose an improved initialization process. Fig. 1 a) illustrates how the full cell was equilibrated in this study. The three electrodes were short-circuited letting the WE and CE equilibrate with the atmosphere, i.e. with $p_{O_2} = 10\ mbar$. The oxygen non-stoichiometry of L2NO4 is decreased down to approximately 0.09 (see half cell section), which corresponds to the discharged state, narrowing the accessible charging capacity to the range [δ$_{min}$ = 0.09 : δ$_{max}$]. To increase the accessible full cell capacity, we propose an alternative method, as illustrated in Fig. 1 b) and c). The cathode is fully oxidized and filled with oxygen (up to the maximal $p_{O_2,eff}$ that the electrode and the capping layer can sustain), and in a second step, the anode is fully reduced. This can be controlled by applying a positive or negative DC voltage between the WE or CE, respectively, and the RE. This corresponds to an initial charging cycle via the reference and loads the optimal amount of oxygen into the OIB. Thus, it allows to cycle the cell over a wider δ range and maximizes the full cell capacity. The corresponding ideal equilibrium of the discharged electrodes is now independent of the atmosphere.

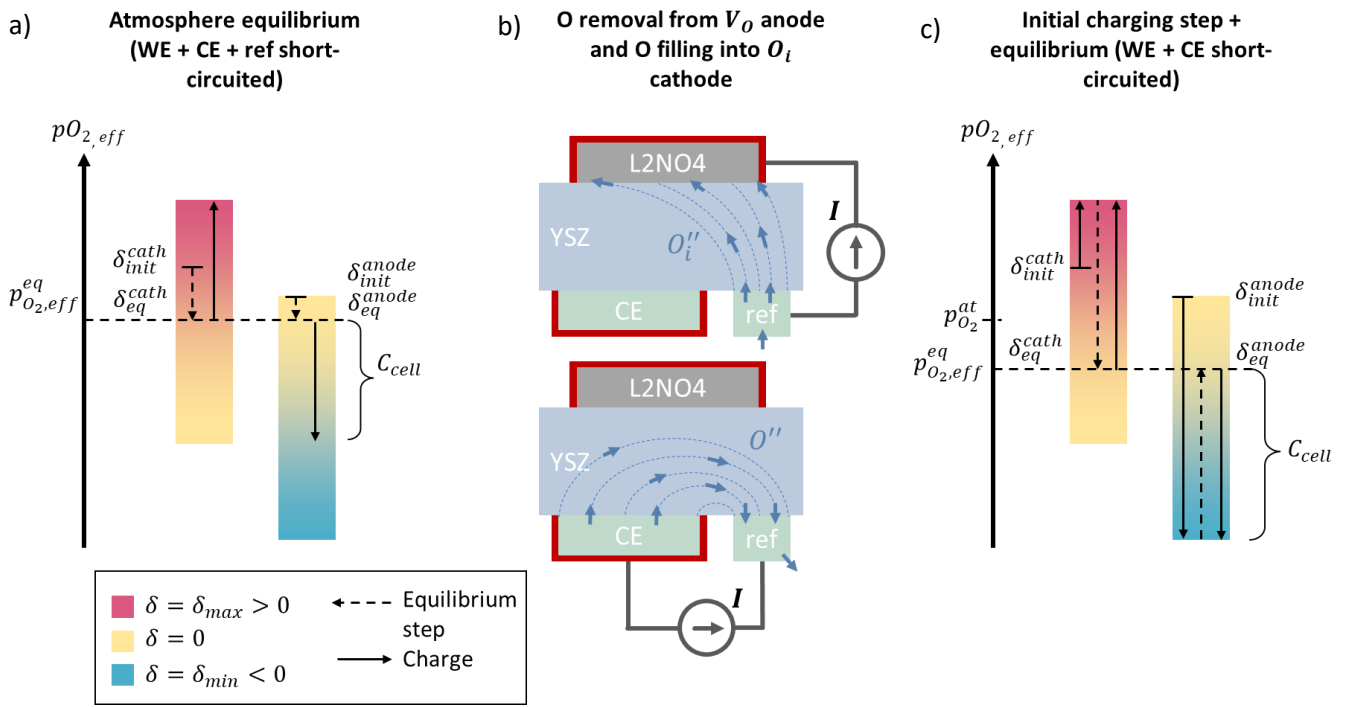

Fig. 1: Schematic illustrations of a) the $p_{O_2,eff}$ equilibrium with the atmosphere (upon shirt-circuiting the electrodes with the reference), and c) the ideal $p_{O_2,eff}$ equilibrium achieved via an initial 2-step charging process, i.e., oxygen filling into the cathode and oxygen removal from the anode shown in b).. This initiation procedure loads the optimum amount of oxygen into the electrodes and therefore maximises the full cell capacity, $C_{cell}$, whereas in a) $C_{cell}$ depends on the atmosphere and excess oxygen inside the OIB undermines the use of the full δ range. The horizontal black dashed lines correspond to the equilibrium state, i.e. the discharged state. The schematics b) show how to fill the cathode with interstitial oxygen (up) and how to create oxygen vacancies in the anode (down).

## Chemical capacitance of L2NO4 with nano-columnar microstructure

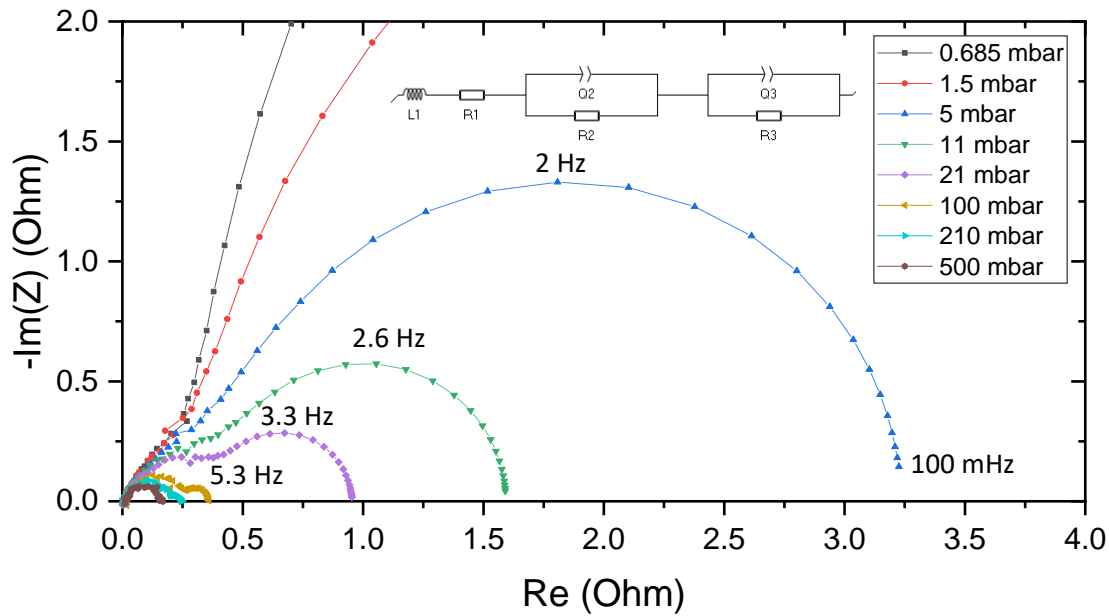

*Fig. S6: Impedance spectra of a symmetrical L2NO4/YSZ/L2NO4 cell using 800 nm nanocolumnar films at 600°C and for various oxygen partial pressures*

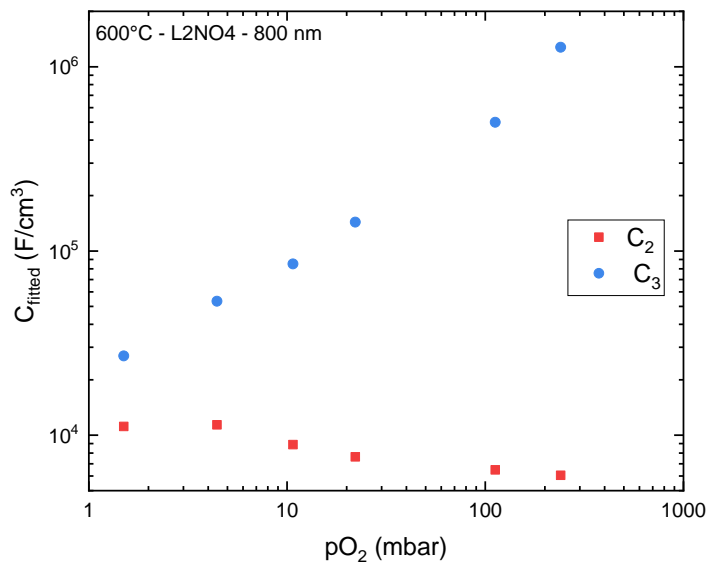

*Fig. S7: Chemical capacitance fitted from impedance spectra. $C_{CPE} = (R^{1-\alpha} \cdot Q)^{\frac{1}{\alpha}}$, where $C_{CPE}$ is the equivalent capacitance of the constant phase element (CPE) in F, noted Q, R is the resistance and α parameter is used to quantify the non-ideal behaviour of the element. The chemical capacitance is calculated from $C_{chem} = \frac{C_{CPE}}{2A \cdot L}$ where A is the active surface of the electrode and L is the thickness of the film. The coefficient 2 comes from the symmetrical cell configuration.*

The resulting values for $C_3$ are in the range $2.10^4$ to $1.10^6\ F/cm^3$, whereas $C_2$ are in the range $5.10^3$ to $1.10^4\ F/cm^3$. In their study, Kim *et al.* found a chemical capacitance of $3.10^4\ F/cm^3$ for a polycrystalline L2NO4 300 nm thick film at 578 °C and 0.01 atm [2].

**Comparison of results for the three L2NO4 microstructures**

*Table 1: Summary of maximum capacity and coulomb efficiency values of the L2NO4/YSZ/LSF half-cell obtained at various temperature, current density and upper cutoff voltages*

| Electrode microstructure | Temperature (°C) | Current density ($\mu A.cm^{-2}$) | Cutoff voltage (V vs RE) | | Max capacity ($mA.h.cm^{-3}$) | Coulomb efficiency (%) |
|---|---|---|---|---|---|---|
| | | | Lower | Upper | | |
| Dense | 350 | 3.6 | 0 | 0.15 | 53.7 | 99.9 |
| Dense | 350 | 8.9 | 0 | 0.15 | 112.7 | 99.3 |
| Dense | 350 | 17.8 | 0 | 0.15 | 161.9 | 96.8 |
| Dense | 350 | 35.6 | 0 | 0.15 | 207.8 | 94.9 |
| Dense | 400 | 17.8 | 0 | 0.15 | 62.8 | 99.9 |
| Dense | 400 | 17.8 | 0 | 0.2 | 319.4 | 93.9 |
| Nano-columnar | 350 | 3.6 | 0 | 0.1 | 56.4 | 99.9 |
| Nano-columnar | 400 | 3.6 | 0 | 0.1 | 21.1 | 99.9 |
| Nano-columnar | 400 | 8.9 | 0 | 0.1 | 33.8 | 99.9 |
| Closed pores | 350 | 8.9 | 0 | 0.1 | 44.1 | 99.6 |
| Closed pores | 400 | 8.9 | 0 | 0.1 | 23.5 | 98.3 |
| Closed pores | 400 | 17.8 | 0 | 0.1 | 29.1 | 99.9 |
| Closed pores | 400 | 35.6 | 0 | 0.15 | 50.5 | 98.4 |
| Closed pores | 400 | 35.6 | 0 | 0.25 | 86.4 | 90.6 |